\documentclass[aps,prd,onecolumn,groupedaddress,showpacs,nofootinbib,amssymb
]{revtex4}
\usepackage[dvips]{graphicx}
\usepackage{amssymb}
\usepackage{amsmath}
\usepackage{epstopdf}
\usepackage{graphicx,,color}
\usepackage{amsfonts}
\usepackage{bm}
\usepackage{cancel}
\usepackage{comment}

\allowdisplaybreaks[4]

\begin{document}

\tolerance=5000



\title{Swampland criteria and neutrino generation in a non-cold dark matter universe}

\author{
Martiros Khurshudyan$^{1}$\thanks{Email: khurshudyan@ice.csic.es}}

\affiliation{
$^1$ Consejo Superior de Investigaciones Cient\'{\i}ficas, ICE/CSIC-IEEC,
Campus UAB, Carrer de Can Magrans s/n, 08193 Bellaterra (Barcelona) Spain \\
}

\begin{abstract}
In this paper, the implications of string Swampland criteria for a dark energy-dominated universe, where we have a deviation from the cold dark matter model, will be discussed. In particular, we have considered two models. One of them is one parameter model, while the second one has been crafted to reveal the dynamics in the deviation. The analysis has been obtained through the use of Gaussian processes (GPs) and $H(z)$ expansion rate data (a $30$-point sample deduced from a differential age method and a $10$-point sample obtained from the radial BAO method). We learned that the tension with the Swampland criteria still will survive as in the cases of the models where dark matter is cold. In the analysis besides mentioned $40$-point $H(z)$ data, we used the latest values of $H_{0}$ reported by the Planck and Hubble missions to reveal possible solutions for the $H_{0}$ tension problem. Finally, the constraints on the neutrino generation number have been obtained revealing interesting results to be discussed yet. This and various related questions have been left to be discussed in forthcoming papers.


\end{abstract}


\maketitle

\section{Introduction}\label{sec:INT}

The discovery of the accelerated expansion of the universe significantly changed our understanding of the universe~\cite{Aghanim_H0} - \cite{Ade2}. It brought new knowledge and imposed new tasks to be solved yet and modification of General Relativity (GR) is one of the direct outcomes of this. On the other hand, quantum corrections also have a central role in crafting viable and advanced modified theories of gravity \cite{Nojiri}~-~\cite{Cognola}~(to mention a few). By modifying GR we search to find an effective way to deal with dark energy, dark matter, inflation, and other relevant problems \cite{Weinberg} - \cite{Mk18} and references therein. Moreover, in light of the most common view GR can not be the ultimate theory of the universe, allowing the modification of GR to be one of the most discussed topics in recent literature. But we can still assume that yet unknown high-energy UV-complete theory can be reduced to GR (low-energy limit). The string theory plays the role of the mentioned unknown UV-complete theory. On the other hand, here we have an interesting situation related to the dS vacua. In particular, until now no dS vacuum has been constructed, owing to numerous problems~\cite{Kachru}~-~\cite{Danielsson}. We took this as a hint indicating that in a consistent quantum theory of gravity, dS does not exist. The mentioned problem allows forming of the Swampland region where inconsistent semi-classical effective field theories should inhabit. On the other hand, we have a Landscape provided by string theory where a vast range of choices fitting our universe in a consistent quantum theory of gravity exists. Therefore, it is not excluded that dS vacua may reside in the Swampland~\cite{Ooguri}~-~\cite{Obied}. Recently two Swampland criteria have been proposed 
\begin{enumerate}
\item $SC1$: The scalar field net excursion in reduced Planck units should satisfy the bound~\cite{Ooguri}
\begin{equation}\label{eq:SC1}
\frac{|\Delta \phi|}{M_{P}} < \Delta \sim O(1),
\end{equation}
\item $SC2$: The gradient of the scalar field potential is bounded by~\cite{Obied}
\begin{equation}\label{eq:SC2}
M_{P}\frac{|V^{\prime}|}{V} > c \sim O(1),
\end{equation}
or \cite{SW_3}
\begin{equation}\label{eq:SC2_new}
M^{2}_{P} \frac{V^{\prime\prime}}{V} < - c \sim O(1).
\end{equation}
\end{enumerate}  
demanding the field to traverse a larger distance, in order to have the domain where the validity of the effective field theory will be fulfilled. Here, GR in the presence of a quintessence field $\phi$ has been considered to be the effective field theory. In the above-given formulation of the Swampland criteria, both $\Delta$ and $c$ are positive constants of order one, while the prime denotes derivative with respect to the scalar field $\phi$, and $M_{P} = 1/\sqrt{8\pi G}$ is the reduced Planck mass. 

On the other hand, GR in the presence of a quintessence field $\phi$ (dark energy) has been used often to explain the accelerated expansion (including cosmic inflation too) of the universe. In this regard, it is highly reasonable to 1) investigate/understand how the constraints on the dark energy model affect the Swampland constraints, and 2) what are conditions to be satisfied in order not to end up in the Swampland. Different attempts in this direction had been taken already (see \cite{Lavinia} - \cite{Akrami}). Even, it appears that with the GPs, it is possible to study and obtain the constraints on the Swampland criteria in a model-independent way \cite{Elizalde_1}.  In other words, with the GPs, it is possible to study the Swampland criteria for a dark energy model without using any explicit model describing the potential of the quintessence field $\phi$ and using a dark energy model to constrain the parameters of that potential. 

Actually, GP is a machine learning tool intensively used in the recent literature to study various cosmology-related problems. A significant part of those studies indicates that with machine learning we can explore unseen and yet unknown physics of our universe in a more efficient way than can be done with traditional tools. One such problem to be mentioned here is the $H_{0}$ tension problem. Basically, the $H_{0}$ tension problem requires explaining why the Planck CMB data analysis and a local measurement from the Hubble Space Telescope give different values for $H_{0}$. We need to understand why in the $\Lambda$CDM scenario the Planck CMB data analysis gives $H_{0} = 67.4 \pm 0.5$ km/s/Mpc, while local measurements from the Hubble Space Telescope yield $H_{0} = 73.52 \pm 1.62$ km/s/Mpc (see \cite{Aghanim_H0} and \cite{Riess_H0}). Recently by Bayesian machine learning various interesting results have been learned about this problem. In particular, a deviation from cold dark matter has been learned giving a solution to the $H_{0}$ tension problem \cite{Elizalde_H0}.  Moreover, another recent study confirms a deviation from the cold dark matter where the GPs and expansion rate data have been used. Even, a hint has been found that the deviation can have a dynamic nature \cite{Elizalde_H0_recent}. We refer readers to the references of this paper to gain more about the problem and what are the alternative options to solve it. Above mentioned results indicate the existence of new knowledge requiring future analysis using new data and statistical tools.

The goal of this study is twofold given the above-mentioned problems and recently learned new hints. In particular, in this study, we will explore the impact of the deviation from the cold dark matter model on the Swampland tension using GPs. Moreover, we will learn how the Swampland tension and the deviation from the cold dark matter model affect the constraints on the neutrino generation number. To our knowledge, this will be the first work trying to do this using GPs. Given the nature of the problem and the tool we applied, some data-related artificial constraints have been imposed which hopefully can be lifted in the near future using other machine learning tools. However, obtained results are in great agreement with other results intensively discussed in the recent literature indicating that the impact of constraints with high precision can be neglected. 

The paper is organized as follows. In Sect.~\ref{sec:DGP} we present the data and discuss the strategy we follow. In Sect.~\ref{sec:Mod} we will discuss the method. In Sect.~\ref{sec:results} we discuss the results obtained from the reconstruction for various scenarios based on three kernels. Moreover, in our analysis, one of the values of $H_{0}$ has been estimated with the GP method and using high-redshift data for $H(z)$, while the other two are taken to be the values from the Planck~\cite{Aghanim_H0} and Hubble~\cite{Riess_H0} missions, respectively. This strategy has been adopted to make the link between the $H_{0}$ tension problem, Swampland criteria tension, and deviation from the cold dark matter model more transparent. The discussion of our results can be found in Sect.~\ref{sec:Discussion}.

\section{Gaussian Processes and Data}\label{sec:DGP}

\begin{table}[t]
  \centering
    \begin{tabular}{ |  l   l   l  |  l   l  l  | p{2cm} |}
    \hline
$z$ & $H(z)$ & $\sigma_{H}$ & $z$ & $H(z)$ & $\sigma_{H}$ \\
      \hline
$0.070$ & $69$ & $19.6$ & $0.4783$ & $80.9$ & $9$ \\
         
$0.090$ & $69$ & $12$ & $0.480$ & $97$ & $62$ \\
    
$0.120$ & $68.6$ & $26.2$ &  $0.593$ & $104$ & $13$  \\
 
$0.170$ & $83$ & $8$ & $0.680$ & $92$ & $8$  \\
      
$0.179$ & $75$ & $4$ &  $0.781$ & $105$ & $12$ \\
       
$0.199$ & $75$ & $5$ &  $0.875$ & $125$ & $17$ \\
     
$0.200$ & $72.9$ & $29.6$ &  $0.880$ & $90$ & $40$ \\
     
$0.270$ & $77$ & $14$ &  $0.900$ & $117$ & $23$ \\
       
$0.280$ & $88.8$ & $36.6$ &  $1.037$ & $154$ & $20$ \\
      
$0.352$ & $83$ & $14$ & $1.300$ & $168$ & $17$ \\
       
$0.3802$ & $83$ & $13.5$ &  $1.363$ & $160$ & $33.6$ \\
      
$0.400$ & $95$ & $17$ & $1.4307$ & $177$ & $18$ \\

$0.4004$ & $77$ & $10.2$ & $1.530$ & $140$ & $14$ \\
     
$0.4247$ & $87.1$ & $11.1$ & $1.750$ & $202$ & $40$ \\
     
$0.44497$ & $92.8$ & $12.9$ & $1.965$ & $186.5$ & $50.4$ \\

$$ & $$ & $$ & $$ & $$ & $$\\ 

$0.24$ & $79.69$ & $2.65$ & $0.60$ & $87.9$ & $6.1$ \\
$0.35$ & $84.4$ & $7$ &  $0.73$ & $97.3$ & $7.0$ \\
$0.43$ & $86.45$ & $3.68$ &  $2.30$ & $224$ & $8$ \\
$0.44$ & $82.6$ & $7.8$ &  $2.34$ & $222$ & $7$ \\
$0.57$ & $92.4$ & $4.5$ &  $2.36$ & $226$ & $8$ \\ 
          \hline
    \end{tabular}
    \vspace{5mm}
\caption{$H(z)$ and its uncertainty $\sigma_{H}$  in  units of km s$^{-1}$ Mpc$^{-1}$. The upper panel consists of thirty samples deduced from the differential age method. The lower panel corresponds to ten samples obtained from the radial BAO method. The table is according to \cite{HTable_0} - \cite{HTable_13}.}
  \label{tab:Table0}
\end{table}

The GPs have two key ingredients and the goal of this section is to give a very brief presentation of them. Mainly, the goal here is to develop some intuition about the method which is using two-point covariance function $K(x,x^{\prime})$ and a mean $\mu(x)$
to get a continuous realization of

\begin{equation}
\xi(x) \propto GP(\mu(x), K(x,x^{\prime}))
\end{equation}
and uncertainty $\Delta \xi(x)$ to get the posterior $\xi(x) \pm \Delta \xi(x)$. It is formed through a Bayesian iterative process allowing the reconstruction of the function representing the data. Using the data directly the GPs allow us to find a form of the function representing the data. It is possible because we consider the observational data to be a realization of the GPs, too. However, the method does not indicate how to choose the kernel and it should be done manually. In the recent literature, various kernel candidates have been considered and one of them is the squared exponent
\begin{equation}\label{eq:kernel1}
K(x,x^{\prime}) = \sigma^{2}_{f}\exp\left(-\frac{(x-x^{\prime})^{2}}{2l^{2}} \right).
\end{equation}
with $\sigma_{f}$ and $l$ hyperparameters to be determined from the minimization of the GP marginal likelihood. The Cauchy kernel
\begin{equation}\label{eq:kernel2}
K_{C}(x,x^{\prime}) = \sigma^{2}_{f} \left [  \frac{l}{ (x - x^{\prime} )^{2}} + l^{2}\right ],
\end{equation}
and the Matern ($\nu = 9/2$)  kernel
$$K_{M}(x,x^{\prime}) = \sigma^{2}_{f} \exp \left(-\frac{3|x-x^{\prime}|}{l} \right) $$
\begin{equation}\label{eq:kernel3}
\times \left[ 1+ \frac{3 |x-x^{\prime}|}{l} + \frac{27(x-x^{\prime})}{7l^{2}} + \frac{18|x-x^{\prime}|^{3}}{7l^{3}} + \frac{27 (x-x^{\prime})^{4}}{35 l^{4}}\right],
\end{equation}
are among the intensively used ones, too.  The $l$ parameter in the above three equations represents the correlation length along which the successive $\xi(x)$ values are correlated. The $\sigma_{f}$ parameter, on the other hand, is used to control the variation in $\xi(x)$ relative to the mean of the process. The readers can follow \cite{Elizalde_H0} - \cite{Elizalde_0} and references therein for a better understanding of the key aspects of the approach and how it can be used in cosmology. The dataset we have used can be found in Table~\ref{tab:Table0}. We use $30$-point samples of $H(z)$ deduced from the differential age method. Then, we add $10$-point samples obtained from the radial BAO method. This allowed us to have good data up to $z=2$ and extend the data range up to $z = 2.4$ where still the value of the Hubble parameter at $z=0$ has not been taken into account. In this study, we will consider three different values for $H_{0}$. In particular, first of all, we allowed the process to guess what the $H_{0}$ should be using the data given in Table~\ref{tab:Table0}. Then, in the other two scenarios, $H_{0} = 67.66 \pm 0.42$ and $H_{0} = 73.52 \pm 1.62$ reported from the Planck and Hubble missions, respectively, have been merged to the available data given in Table~\ref{tab:Table0}. Therefore, we could craft two datasets with two different $H_{0}$ values to be used in the reconstruction process. To save space we refer to the upper part of Table~\ref{tab:Table1} to find more about the estimated $H_{0}$ value for considered kernels \footnote{We have used the GaPP~(Gaussian Processes in Python) package developed by Seikel et al \cite{Seikel}.}.

\begin{figure}[h!]
 \begin{center}$
 \begin{array}{cccc}
\includegraphics[width=150 mm]{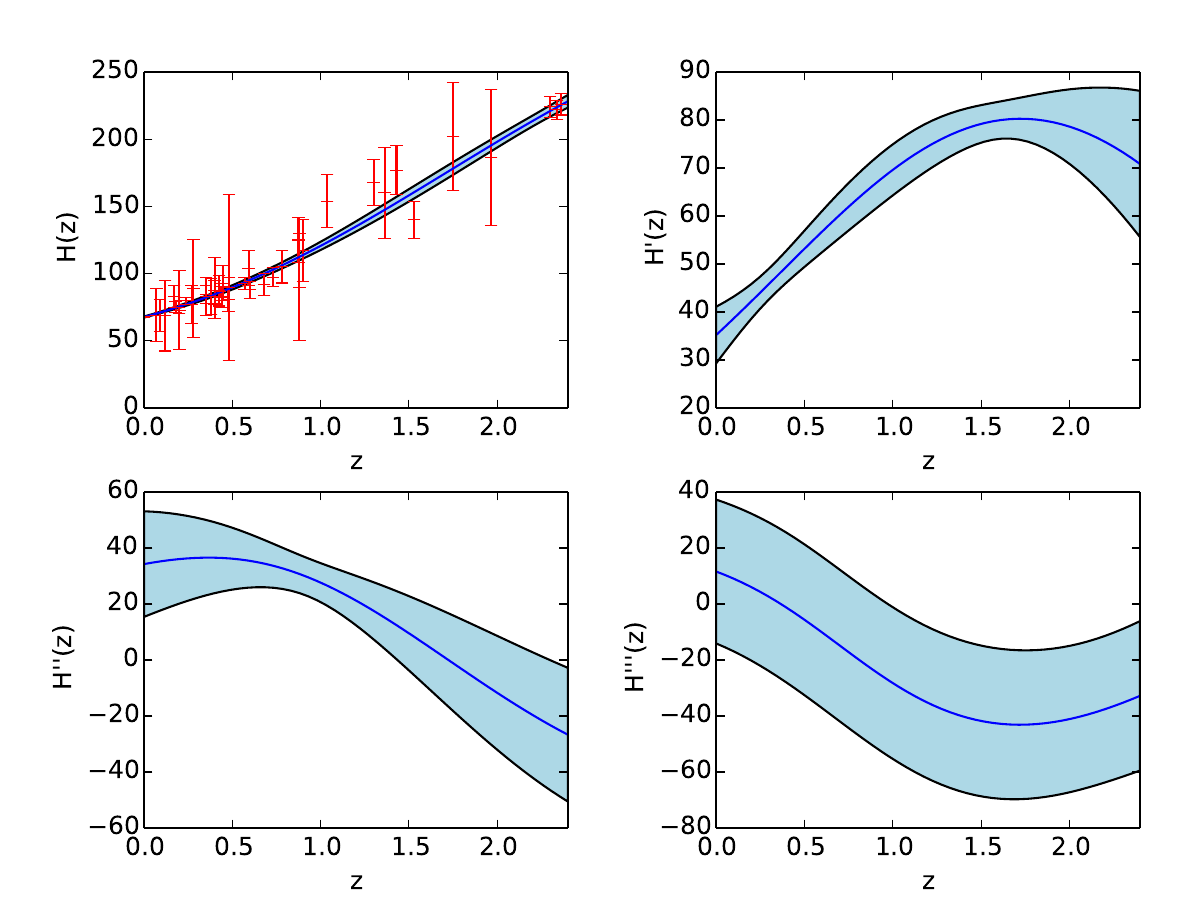} &&
 \end{array}$
 \end{center}
\caption{GP reconstruction of $H(z)$, $H^{\prime}(z)$, $H^{\prime \prime}(z)$, and $H^{\prime \prime \prime}(z)$, for the $40$-point sample deduced from the differential age method, with the additional 10-point sample obtained from the radial BAO method, when $H_{0} = 67.66 \pm 0.42$ reported by the Planck mission. The $^{\prime}$ means derivative with respect to the redshift variable $z$. The kernel is the Matern~($\nu = 9/2$) kernel given by Eq. (\ref{eq:kernel3}).}
 \label{fig:Fig0_1}
\end{figure}

\begin{figure}[h!]
 \begin{center}$
 \begin{array}{cccc}
\includegraphics[width=150 mm]{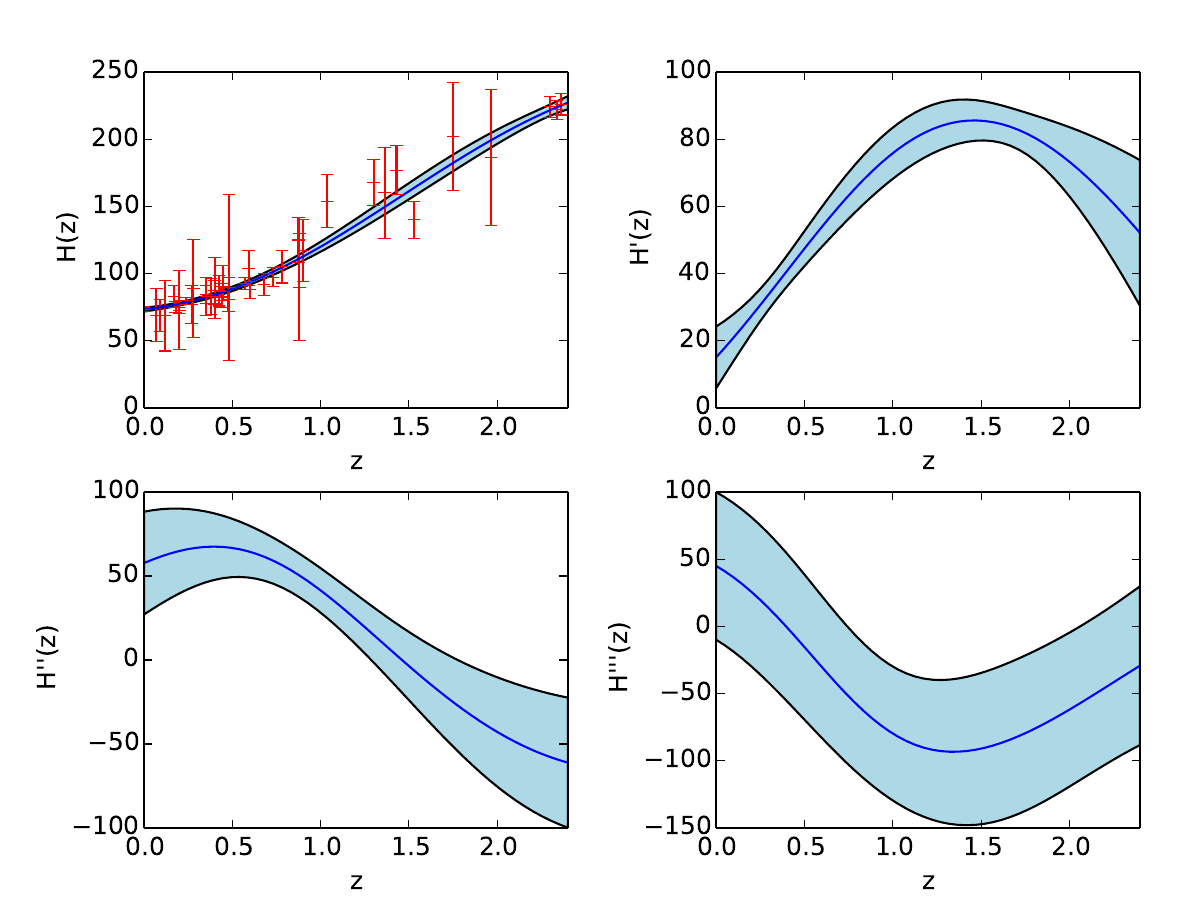} &&
 \end{array}$
 \end{center}
\caption{GP reconstruction of $H(z)$, $H^{\prime}(z)$, $H^{\prime \prime}(z)$ and $H^{\prime \prime \prime}(z)$ for the $40$-point sample deduced from the differential age method, with the additional 10-point sample obtained from the radial BAO method, when $H_{0} = 73.52 \pm 1.62$ reported by the Hubble mission. The $^{\prime}$ means derivative with respect to the redshift $z$. The kernel is the Matern~($\nu = 9/2$) kernel given by Eq. (\ref{eq:kernel3})}
 \label{fig:Fig0_2}
\end{figure}

To end this section we present the reconstructions of $H(z)$ function and its higher order derivatives for $H_{0} = 67.66 \pm 0.42$ and $H_{0} = 73.52 \pm 1.62$ for the Matern~($\nu = 9/2$) kernel, Eq. (\ref{eq:kernel3}). They can be found in Fig. \ref{fig:Fig0_1}  and Fig.~\ref{fig:Fig0_2}, respectively. The reconstruction of the other cases used in this paper can be found in the references of this paper and they are not presented here only to save our place. 
 
In the next section, we will revise how we can estimate the bounds of $SC2$ in a model-independent way and how to infer the constraints on the neutrino number generation when we have certain deviations from the cold dark matter model.

\section{Model}\label{sec:Mod}

Here we consider GR in the presence of a quintessence field $\phi$ to be the effective field theory described by the following action (see for instance \cite{Elizalde_1})
\begin{equation}
S = \int d^{4} x \sqrt{-g} \left( \frac{M^{2}_{P}}{2} R - \frac{1}{2} \partial_{\mu} \phi \partial^{\mu} \phi -V(\phi) \right ) + S_{m},
\end{equation}
where $S_{m}$ corresponds to the matter, $M_{P} = 1/\sqrt{8\pi G}$ is the reduced Planck mass, $R$ is the Ricci scalar and $V(\phi)$ is the field potential. According to this scenario when we consider the FRWL universe, the dynamics of the scalar field's dark energy and matter will be described by the following equations 
\begin{equation}\label{eq:drhoPi}
\dot{\rho}_{\phi} + 3 H (\rho_{\phi} + P_{\phi}) = 0,
\end{equation}
\begin{equation}\label{eq:drhoDm}
\dot{\rho}_{dm} + 3 H (\rho_{m} + P_{m} )= 0,
\end{equation}
where $\rho_{\phi}$ and $P_{\phi}$ are the energy density and pressure describing the quintessence field $\phi$, while $\rho_{m}$ and $P_{m}$ are the energy density and pressure of the matter, respectively. We know also that they are related to each other through the Friedmann equations, as follows
\begin{equation}\label{eq:F1}
H^{2} = \frac{1}{3} (\rho_{\phi} + \rho_{m}), 
\end{equation}
and 
\begin{equation}\label{eq:F2}
\dot{H} + H = -\frac{1}{6} (\rho_{\phi} + \rho_{m} + 3 ( P_{\phi} + P_{m})),
\end{equation}
with $H = \dot{a}/a$ to be the Hubble parameter. On the other hand, for the spatially homogeneous scalar field, the energy density and pressure are of the following form
\begin{equation}\label{eq:rhoPhi}
\rho_{\phi} = \frac{1}{2} \dot{\phi} + V(\phi),
\end{equation}
and
\begin{equation}\label{eq:pPhi}
P_{\phi} = \frac{1}{2} \dot{\phi} - V(\phi),
\end{equation}
where the dot is the derivative w.r.t to the cosmic time. But it is easy to see that from Eqs.~(\ref{eq:rhoPhi}) and~(\ref{eq:pPhi}) we have 

\begin{equation}\label{eq:dphi2}
\dot{\phi}^{2} = \rho_{\phi} + P_{\phi},
\end{equation} 
while
\begin{equation}\label{eq:Vphi}
V(\phi) = \frac{\rho_{\phi} - P_{\phi}}{2}.
\end{equation}
    
After the discussion given above, it is easy to see that we can use Eqs.~(\ref{eq:F1}) and ~(\ref{eq:F2}), and the model of the matter to constrain Swampland criteria, Eq. (\ref{eq:SC1}) and Eq. (\ref{eq:SC2}) (or  Eq. (\ref{eq:SC2_new})), respectively. Because  Eq. (\ref{eq:SC1}) and Eq. (\ref{eq:SC2}) (or  Eq. (\ref{eq:SC2_new})), can be expressed in terms of the Hubble function and its higher order derivatives which will be reconstructed using GPs from the expansion rate data. In other words, we are in the position to reconstruct $SC2$ and estimate its upper bound in a model-independent way using directly observational data. Coming back to the form of $SC2$ we need take into account that $d V(\phi)/d\phi = (dV/dz)/(d\phi/dz)$, where $d\phi/dz$ should be calculated from Eq.~(\ref{eq:dphi2}), and that $\dot{\phi} = -(1+z)H \phi^{\prime}$. The results of the study, for the strategies discussed in Sect.~\ref{sec:DGP} are presented in the next section.  

To end this section, we need to define the last element required to perform the analysis and it is the matter model. In particular, we consider two models where the matter part differs from the cold dark matter described by $P_{dm} = \omega_{dm} \rho_{dm}$ equation. As the first model, we consider the case when $\omega_{dm}$ is a non-zero constant, while we consider $\omega_{dm} = \omega_{0} + \omega_{1} z$ to be the second model, respectively. In both models $\omega_{0}$ and $\omega_{1}$ are the model parameters to be learned. Moreover, in our analysis, we included neutrinos through radiation
\begin{equation}
\rho_{r} = 3 H^{2}_{0} \Omega^{(0)}_{r} (1+z)^{4},
\end{equation}
where 
\begin{equation}
\Omega^{(0)}_{r} = \Omega^{(0)}_{\gamma} (1 + 0.68 (N_{eff} /3) )
\end{equation}
following the very well-known strategy used in modern cosmology, with $\Omega^{(0)}_{\gamma}$ and $N_{eff}$ free parameters to be constrained too. Here $N_{eff}$ is the neutrino generation number.

\section{The reconstruction Results}\label{sec:results}

In this section, we present and discuss the key aspects we have inferred from the reconstruction using the GP. The starting point in our analysis is based on the assumptions we have used for $\rho_{m}$  to define the energy density of the scalar field, $\rho_{\phi}$, Eq.~(\ref{eq:F1}). In this paper, we have considered two models only, but the discussion given in the previous section is applicable to any other model describing deviations from the cold dark matter model. Obtained results indicate a need to continue research in this direction involving new data and tools to reveal the true nature of learned phenomena and their impact on various fundamental problems of modern cosmology. We need to stress that for the dark energy-dominated universe the first Swampland criteria, Eq. (\ref{eq:SC1}) is always satisfied therefore we will not present the reconstruction of it in the discussion below. 

\subsection{Model with $\omega_{dm} = \omega_{0}$}

The first model we consider is the model where the dark matter equation is $P_{dm} = \omega_{0} \rho_{dm}$ with $\omega_{0} \neq 0$ to be learned. The dynamics of the energy density of this model according to Eq.~(\ref{eq:drhoDm}) will be given by the following equation \cite{Elizalde_H0}
\begin{equation}\label{eq:NCDM_1}
\rho_{dm} =   3 H^{2}_{0} \Omega^{(0)}_{dm} (1 + z)^{3 (1 + \omega_{0} )} ,
\end{equation}
where $H_{0}$ and $\Omega^{(0)}_{dm}$ are the Hubble parameter and the fraction of the dark matter at $z=0$, respectively. Therefore, for $\rho_{de}$ ($\rho_{\phi}$) we will have (see  Eq. (\ref{eq:F1}))
\begin{equation}\label{eq:NCDM_1_rhode}
\rho_{de} = 3 H^{2} - \rho_{dm} - \rho_{r}= 3H^{2} - 3 H^{2}_{0} \Omega^{(0)}_{dm} (1 + z)^{3 (1 + \omega_{0} )} - 3 H^{2}_{0} \Omega^{(0)}_{r} (1+z)^{4}.
\end{equation}
Using the above-given equation and some simple algebra, eventually for $\rho^{\prime}_{de} = d \rho_{de}/dz$ we will get
\begin{equation}
\rho^{\prime}_{de} = 6 H H^{\prime} -9 H^2_{0} (1 + \omega_{0}) \Omega^{(0)}_{dm} (1 + z)^{3 \omega_{0}+2} - 12 H^{2}_{0} \Omega^{(0)}_{r} (1+z)^{3}.
\end{equation}
It is not hard to see that after some algebra for $P_{de}$ ($P_{\phi}$) we will get
\begin{equation}\label{eq:phi}
P_{de} = -3 H^2 - 3 H^{2}_{0} \omega_{0} \Omega^{(0)}_{dm} (1+ z)^{3 (1 + \omega_{0} )} - H^{2}_{0} \Omega^{(0)}_{r} (1+z)^{4} + 2 (1 + z) H H^{\prime} .
\end{equation}
On the other hand, using Eq.~(\ref{eq:phi}) we can calculate  $dP_{de}/dz$ used to estimate $SC2$, Eq. (\ref{eq:SC2}). The constraints on the parameters for this case we have obtained can be found in Table \ref{tab:Table1}. From where we see that when the $H_{0}$ has been estimated using available expansion rate data we got a model of the universe where $\Omega^{(0)}_{dm} \approx 0.262$ according to the mean of the reconstruction when the kernel is given by Eq. (\ref{eq:kernel1}). We estimated $\Omega^{(0)}_{dm} \approx 0.266$ when the kernels are given by Eq. (\ref{eq:kernel2}) and Eq. (\ref{eq:kernel3}), respectively. In all three cases, the deviation from the cold dark matter model has been learned (see the upper panel of Table \ref{tab:Table1}). On the other hand, when $H_{0} = 73.52 \pm 1.62$ km/s/Mpc has been merged to the expansion rate data and used in the reconstruction we estimated that $\Omega^{(0)}_{dm} \approx 0.273$ when the kernel is given by Eq. (\ref{eq:kernel1}). On the other hand, when the kernel has been given by Eq. (\ref{eq:kernel2}) we estimated  $\Omega^{(0)}_{dm}$ to be about $0.278$ according to the mean of the reconstruction. Finally, we got $\Omega^{(0)}_{dm} \approx 0.281$ when the kernel is given by  Eq. (\ref{eq:kernel3}) indicating a huge impact of the kernel on the estimations of dark matter fraction $\Omega^{(0)}_{dm}$ in our universe. The results corresponding to this case can be found in the middle panel of Table \ref{tab:Table1}). The bottom panel of Table \ref{tab:Table1}) represents the case when $H_{0} = 67.66 \pm 0.42$ km/s/Mpc from the Planck CMB data analysis has been merged together with available expansion rate data given in Table \ref{tab:Table0} and used in the reconstruction. From the obtained results we infer a very important result indicating that to solve the $H_{0}$ tension problem a strong deviation from the cold dark matter model is required. Moreover, this will affect the constraints on $\Omega^{(0)}_{\gamma}$. However, the constraints on $N_{eff}$ indicate that in all cases the neutrino generation number will be three. 

\begin{table}[ht]
	\centering
	
	\begin{tabular}{|c|c|c|c|c|c|c|} 
		\hline
		$Kernel$  & $\Omega^{(0)}_{dm}$ & $H_{0}$ & $\omega_{0}$ & $\Omega^{(0)}_{\gamma}$  & $N_{eff}$\\
		\hline
		 Squared Exponent & $0.262 \pm 0.011$ & $71.286 \pm 3.743$ km/s/Mpc & $-0.071 \pm 0.011$  & $0.00023 \pm 0.00002$ & $2.98 \pm 0.08$\\
		\hline
		Cauchy & $0.266 \pm 0.012$ & $71.472 \pm 3.879$ km/s/Mpc & $-0.075\pm 0.011$  & $0.00022 \pm 0.00002$ & $2.95 \pm0.08$\\
		\hline
		Matern $(\nu = 9/2)$ & $0.266 \pm 0.011$ & $71.119 \pm 3.867$ km/s/Mpc & $-0.076\pm 0.011$ & $0.00021 \pm 0.00002$ & $2.97 \pm0.08$\\
		\hline
     
     \multicolumn{3}{c}{} \\ \hline
     
      		Squared Exponent & $0.273 \pm 0.011$ & $73.52 \pm 1.62$ km/s/Mpc & $-0.075 \pm 0.011$ &  $0.00022 \pm 0.00005$ & $2.94 \pm 0.11$ \\
		\hline
		Cauchy & $0.278 \pm 0.011$ & $73.52 \pm 1.62$ km/s/Mpc & $-0.083\pm 0.012$  & $0.00019 \pm 0.00007$ & $2.92 \pm 0.15$\\
		\hline
		Matern $(\nu = 9/2)$ & $0.281 \pm 0.012$ & $73.52 \pm 1.62$ km/s/Mpc & $-0.088\pm 0.013$ &  $0.00015 \pm 0.00005$ & $2.96 \pm 0.11$\\
		\hline

	\multicolumn{3}{c}{} \\ \hline
	
		Squared Exponent & $0.293 \pm 0.013$ & $67.66 \pm 0.42$ km/s/Mpc & $-0.051 \pm 0.017$ & $0.00015 \pm 0.00005$ & $3.04 \pm 0.12$  \\
		\hline
		Cauchy & $0.291 \pm 0.013$ & $67.66 \pm 0.42$ km/s/Mpc & $-0.045\pm 0.012$ & $0.00013 \pm 0.00002$ & $3.02 \pm 0.13$  \\
		\hline
		Matern $(\nu = 9/2)$ & $0.291 \pm 0.011$ & $67.66 \pm 0.42$ km/s/Mpc & $-0.049\pm 0.015$ & $0.00015 \pm 0.00002$ & $2.95 \pm 0.11$  \\
		\hline

	\end{tabular}
	\caption{Constraints on the parameters for the cosmological model where the deviation from the cold dark matter is given by Eq.~(\ref{eq:NCDM_1}) and $\omega_{0} \neq 0$. The constraints have been obtained for three kernels, Eq. (\ref{eq:kernel1}), Eq. (\ref{eq:kernel2}), and Eq. (\ref{eq:kernel3}), respectively. In particular, the upper part of the table stands for the case when the $H_{0}$ value has been predicted from the GP. The middle part of the table stands for the case when the $H_{0} = 73.52 \pm 1.62$ km/s/Mpc from the Hubble Space Telescope has been merged with available expansion rate data given in Table \ref{tab:Table0} to reconstruct the $H(z)$ and $H^{\prime}(z)$. Finally, the lower part of the table stands for the case when the $H_{0} = 67.4 \pm 0.5$ km/s/Mpc from the Planck CMB data analysis has been merged together with available expansion rate data given in Table \ref{tab:Table0} and used in the reconstruction.}
	\label{tab:Table1}
\end{table}

\begin{figure}[h!]
 \begin{center}$
 \begin{array}{cccc}
 \includegraphics[width=90 mm]{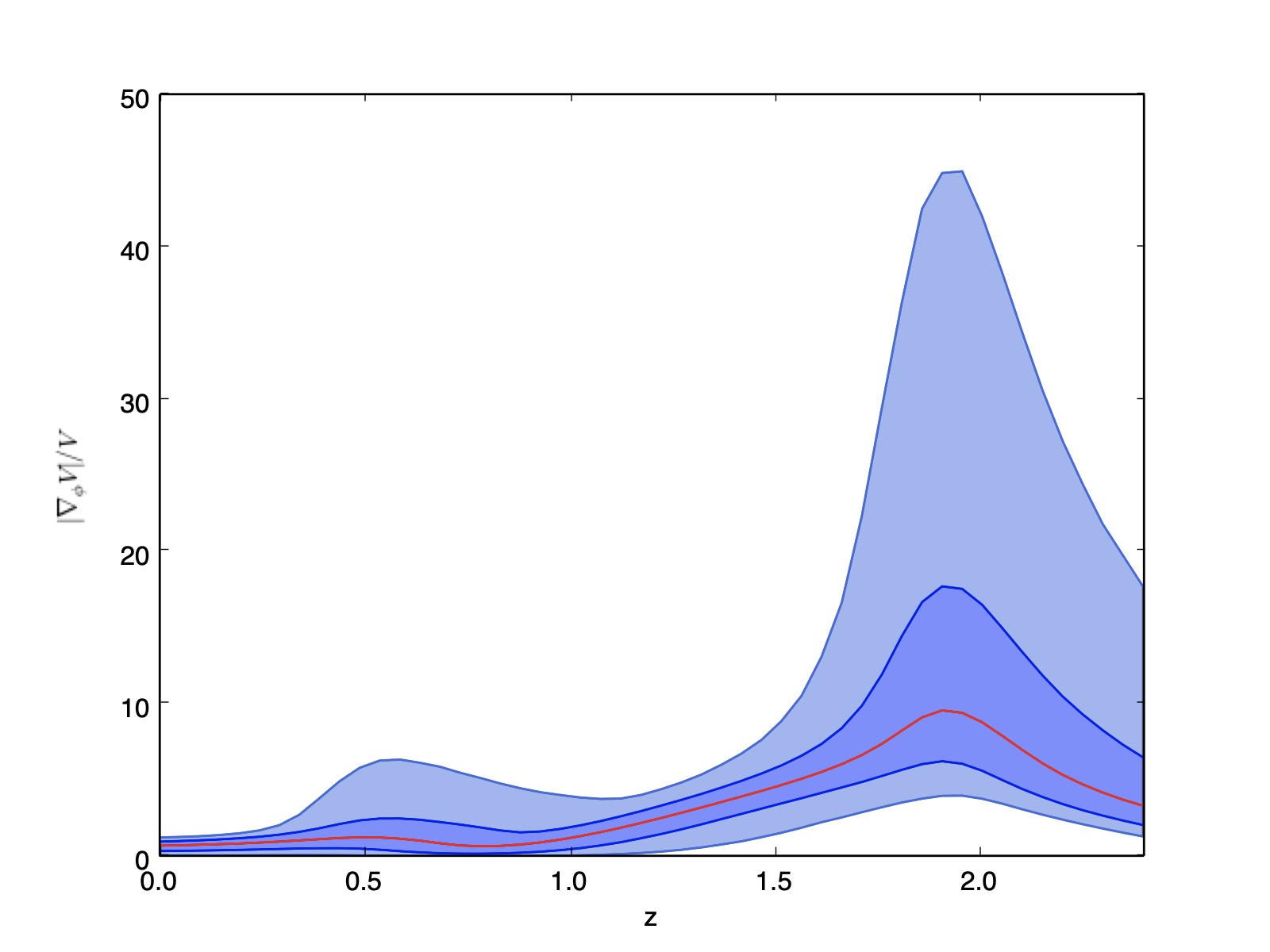} &&
\includegraphics[width=90 mm]{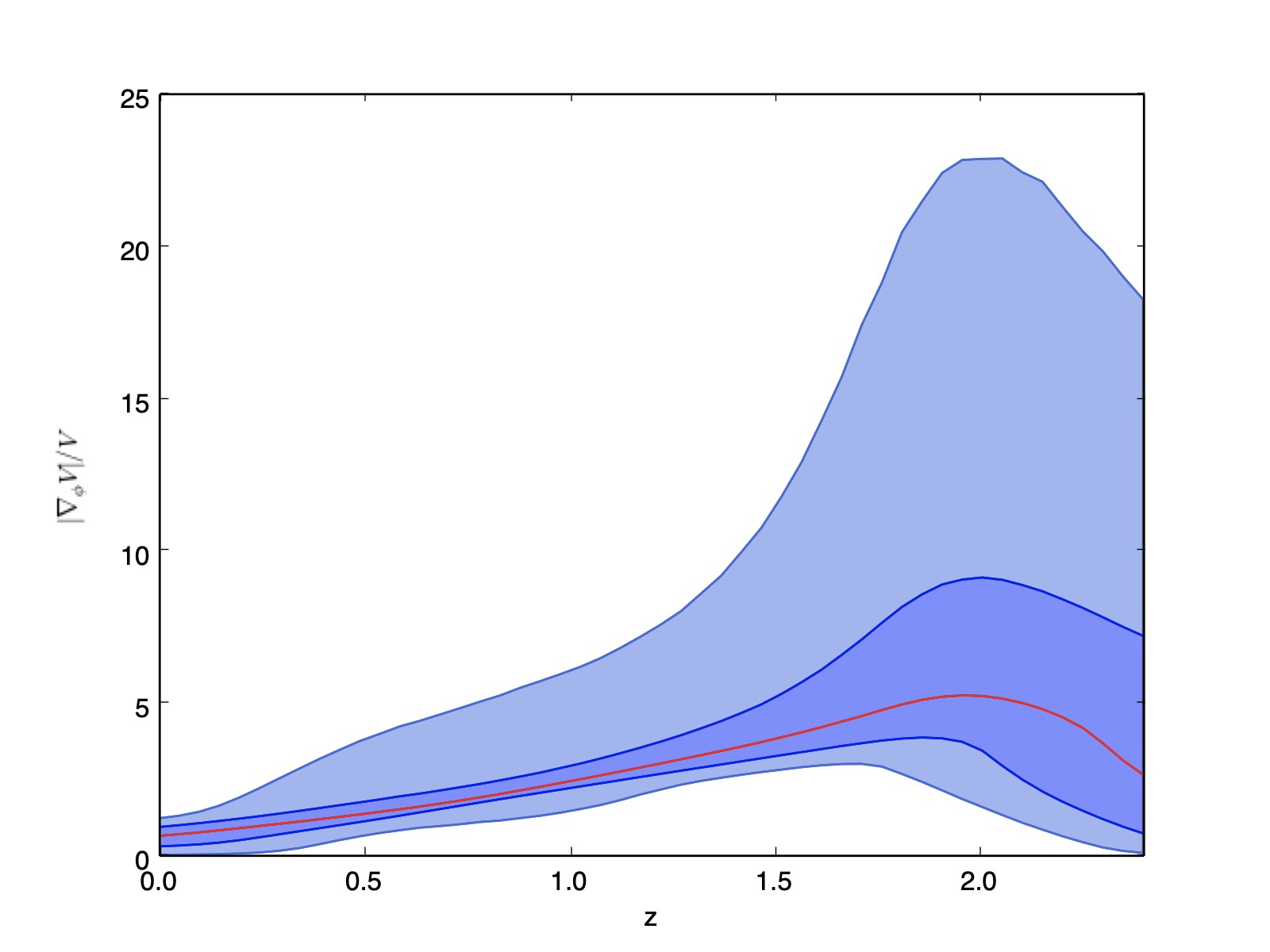} \\
 \end{array}$
 \end{center}
\caption{The left-hand side plot represents the reconstruction of the $|V^{\prime}|/V$, Eq.~(\ref{eq:SC2}), when the $H_{0} = 73.52 \pm 1.62$ km/s/Mpc from the Hubble Space Telescope has been merged with the $H(z)$ data depicted in Table~\ref{tab:Table0} and used in the reconstruction process. The right-hand side plot represents the reconstruction of the $|V^{\prime}|/V$, Eq.~(\ref{eq:SC2}), when the $H_{0} = 67.4 \pm 0.5$ km/s/Mpc from the Planck CMB data analysis has been merged with the $H(z)$ data depicted in Table~\ref{tab:Table0} and used in the reconstruction process. In both cases the squared exponent kernel given by Eq.~(\ref{eq:kernel1}) has been used. The solid line is the mean of the reconstruction and the shaded blue regions are the $68\%$ and $95\%$ C.L. of the reconstruction, respectively. The model is given by by Eq.~(\ref{eq:NCDM_1}) and $\omega_{0} \neq 0$.}
 \label{fig:Fig1}
\end{figure}

The model-independent reconstruction of the $SC2$ given by Eq. (\ref{eq:SC2}) can be found in Fig. \ref{fig:Fig1}. In particular, two plots of Fig. \ref{fig:Fig1} represent the reconstruction when the squared exponent kernel given by Eq.~(\ref{eq:kernel1}) has been used. In the case of the left-hand side plot, we have the reconstruction when the $H_{0} = 73.52 \pm 1.62$ km/s/Mpc from the Hubble Space Telescope has been merged with the $H(z)$ data depicted in Table~\ref{tab:Table0} and used in the reconstruction process. On the other hand, the right-hand side plot represents the reconstruction of the $|V^{\prime}|/V$, Eq.~(\ref{eq:SC2}), when the $H_{0} = 67.4 \pm 0.5$ km/s/Mpc from the Planck CMB data analysis has been merged with the $H(z)$ data and used in the reconstruction process. To save our place we did not present the reconstruction results corresponding to the other kernels. Because in all cases we have obtained similar qualitative results indicating that in the dark energy-dominated universe the recent form of the Swampland criteria is in huge tension with the expansion rate data and future development in this direction is a must. Moreover, we see that the deviation from the cold dark matter model is unable to reduce the tension. Here we confirm that the non-gravitational interaction is also not able to reduce this tension confirming previously obtained results. However, we see clearly that even a theory not residing in Swampland can end up or not end up in Swampland. Moreover, a theory residing in Swampland can end up or not end up in Swampland. In other words, we have four possibilities indicating that inferring a solution for the $H_{0}$ tension problem from the Swampland criteria is also possible but is an extremely hard problem. In other words, the conclusion about the $H_{0}$ tension problem from the Swampland criteria is not unique. The constraints on the $N_{eff}$ inferred from the Swampland criteria indicates that we should expect three generations of neutrinos in our universe even if we have a deviation from the cold dark matter model, Eq. (\ref{eq:NCDM_1}). We need to stress again that the model-independent reconstruction means that we did not use any specific form of scalar field potential and dark energy model to constrain its parameters. 

\subsection{Model with $\omega_{m} = \omega_{0} + \omega_{1} z$}

The second model we consider here is another simple model that can reveal possible dynamics in the deviations from the cold dark matter model. The linear model we have crafted is given below \cite{Elizalde_H0_recent}
\begin{equation}\label{eq:omega_2}
\omega_{m} = \omega_{0} + \omega_{1} z
\end{equation}
providing the dynamics of dark matter to be
\begin{equation}\label{eq:NCDM_2}
\rho_{dm} =   3 H^{2}_{0} \Omega^{(0)}_{dm} e^{3 \omega_{1} z} (1+ z)^{3 (1 + \omega_{0}-\omega_{1} )},
\end{equation}
because $P_{dm} = (\omega_{0} + \omega_{1} z) \rho_{dm}$, where $\omega_{0}$ and $\omega_{1}$ are the free parameters to be learned. It is easy to see that in this case
\begin{equation}\label{eq:drhodm_2}
\frac{d\rho_{dm}}{dz} = 9  H^{2}_{0} \Omega_{dm} e^{3 \omega_{1} z} (1 + z)^{3 \omega_{0}-3 \omega_{1}+2} (1 + \omega_{0}+\omega_{1} z),
\end{equation}
\begin{equation}\label{eq:rhode_1}
\rho_{de} = 3 H^{2} - \rho_{dm} = 3 H^{2} - 3 H^{2}_{0} \Omega^{(0)}_{dm} e^{3 \omega_{1} z} (1+ z)^{3 (1 + \omega_{0}-\omega_{1} )} - 3 H^{2}_{0} \Omega^{(0)}_{r} (1+z)^{4},
\end{equation}
and
\begin{equation}\label{eq:drhode_2}
\frac{d\rho_{de}}{dz} = 6 H H^{\prime} -9 H^{2}_{0} \Omega^{(0)}_{dm} e^{3 \omega_{1} z} (1 + z)^{3 \omega_{0}-3 \omega_{1}+2} (1 + \omega_{0}+\omega_{1} z) - 12 H^{2}_{0} \Omega^{(0)}_{r} (1+z)^{3},
\end{equation}
respectively. Therefore, after some algebra for the dark energy, we will have
\begin{equation}\label{eq:NCDM_2_omegade}
\omega_{de} = - \frac{3 H^{2}  + 3 H^{2}_{0} \Omega^{(0)}_{dm} e^{3 \omega_{1} z} (\omega_{0}+\omega_{1} z) (1 + z)^{3 (1 + \omega_{0}- \omega_{1} )} +  H^{2}_{0} \Omega^{(0)}_{r} (1+z)^{4}-  2 (1 + z) H H^{\prime} }{3 H^{2} - 3 H^{2}_{0} \Omega^{(0)}_{dm} e^{3 \omega_{1} z} (1 + z)^{3 (1 + \omega_{0} - \omega_{1} ) } - 3 H^{2}_{0} \Omega^{(0)}_{r} (1+z)^{4}},
\end{equation}
allowing to reconstruct the equation of state dynamics for dark energy since for the pressure $P_{de}$ we have
\begin{equation}
P_{de} = -3 H^{2} - 3 H^{2}_{0} \Omega^{(0)}_{dm} e^{3 \omega_{1} z} (\omega_{0}+\omega_{1} z) (1 + z)^{3 (1 + \omega_{0} -\omega_{1})} - H^{2}_{0} \Omega^{(0)}_{r} (1+z)^{4}+ 2 (1 + z) H H^{\prime} .
\end{equation}

The constraints on $\Omega^{(0)}_{dm}$, $\omega_{0}$ and $\omega_{1}$ we obtained can be found in Table \ref{tab:Table2}. The upper part of Table \ref{tab:Table2} represents the constraints when the $H_{0}$ has been estimated using only the expansion rate data given in Table \ref{tab:Table0}. The middle part of Table \ref{tab:Table2} represents the constraints when we merged the $H_{0} = 73.52 \pm 1.62$ km/s/Mpc with the available $H(z)$ data. While the constraints we have obtained when the $H_{0} = 67.4 \pm 0.5$ km/s/Mpc with the available $H(z)$ data has been merged, can be found in the lower part of Table \ref{tab:Table2}. A closer look at obtained results indicates that the deviation from the cold dark matter solving the $H_{0}$ tension problem is not able to reduce the existing tension with the Swampland criteria. 

\begin{table}[ht]
	\centering
	
	\begin{tabular}{|c|c|c|c|c|c|c|c|} 
		\hline
		$Kernel$  & $\Omega^{(0)}_{dm}$  & $\omega_{0}$ & $\omega_{1}$  & $\Omega^{(0)}_{\gamma}$  & $N_{eff}$\\
		\hline
		 Squared Exponent & $0.267 \pm 0.011$ &  $-0.074 \pm 0.011$  & $-0.015 \pm 0.003$ & $0.00021 \pm 0.00002$ & $2.89 \pm 0.09$\\
		\hline
		Cauchy & $0.263 \pm 0.011$ & $-0.074\pm 0.011$ & $-0.0015 \pm 0.011$ & $0.00021 \pm 0.00002$ & $2.89 \pm 0.09$ \\
		\hline
		Matern $(\nu = 9/2)$ & $0.267 \pm 0.011$ & $-0.074\pm 0.011$ & $-0.0027 \pm 0.0015$ & $0.00019 \pm 0.00002$ & $2.96 \pm 0.03$\\
		\hline
     
     \multicolumn{4}{c}{} \\ \hline
     
      		Squared Exponent & $0.272 \pm 0.012$ & $-0.065 \pm 0.012$  & $-0.014 \pm 0.011$ & $ 0.00017 \pm 0.00002$ & $2.99 \pm 0.03$\\
		\hline
		Cauchy & $0.271 \pm 0.011$ & $-0.065\pm 0.011$ & $-0.009 \pm 0.005$ & $0.00017 \pm 0.00002$ & $2.99 \pm 0.05$\\
		\hline
		Matern $(\nu = 9/2)$ & $0.271 \pm 0.012$ & $-0.068\pm 0.012$ & $-0.009 \pm 0.007$ & $0.00021 \pm 0.00002$ & $2.91 \pm 0.08$\\
		\hline

	\multicolumn{4}{c}{} \\ \hline
	
		Squared Exponent & $0.271 \pm 0.013$ & $-0.022 \pm 0.014$  & $-0.009 \pm 0.007$ & $0.00021 \pm 0.00003$ & $ 2.94 \pm 0.05$\\
		\hline
		Cauchy & $0.273 \pm 0.013$ & $-0.028 \pm 0.017$ & $-0.008 \pm 0.007$ & $0.00019 \pm 0.00002$ & $2.85 \pm 0.04$\\
		\hline
		Matern $(\nu = 9/2)$ & $0.273 \pm 0.012$ & $-0.027 \pm 0.015$ & $-0.009 \pm 0.007$ & $0.00018 \pm 0.00002$ & $2.91 \pm 0.03$\\
		\hline

	\end{tabular}
	\caption{Constraints on the parameters for the cosmological model where the deviation from the cold dark matter is described by Eq.~(\ref{eq:omega_2}) and Eq.~(\ref{eq:NCDM_2}), respectively. The constraints have been obtained for three kernels, Eq. (\ref{eq:kernel1}), Eq. (\ref{eq:kernel2}), and Eq. (\ref{eq:kernel3}), respectively. In particular, the upper part of the table stands for the case when the $H_{0}$ value has been predicted from the GP. The middle part of the table stands for the case when the $H_{0} = 73.52 \pm 1.62$ km/s/Mpc from the Hubble Space Telescope has been merged together with available expansion rate data given in Table \ref{tab:Table0}. Finally, the lower part of the table stands for the case when the $H_{0} = 67.4 \pm 0.5$ km/s/Mpc from the Planck CMB data analysis has been merged together with available expansion rate data given in Table \ref{tab:Table0} and used in the reconstruction.}
	\label{tab:Table2}
\end{table}

\begin{figure}[h!]
 \begin{center}$
 \begin{array}{cccc}
 \includegraphics[width=90 mm]{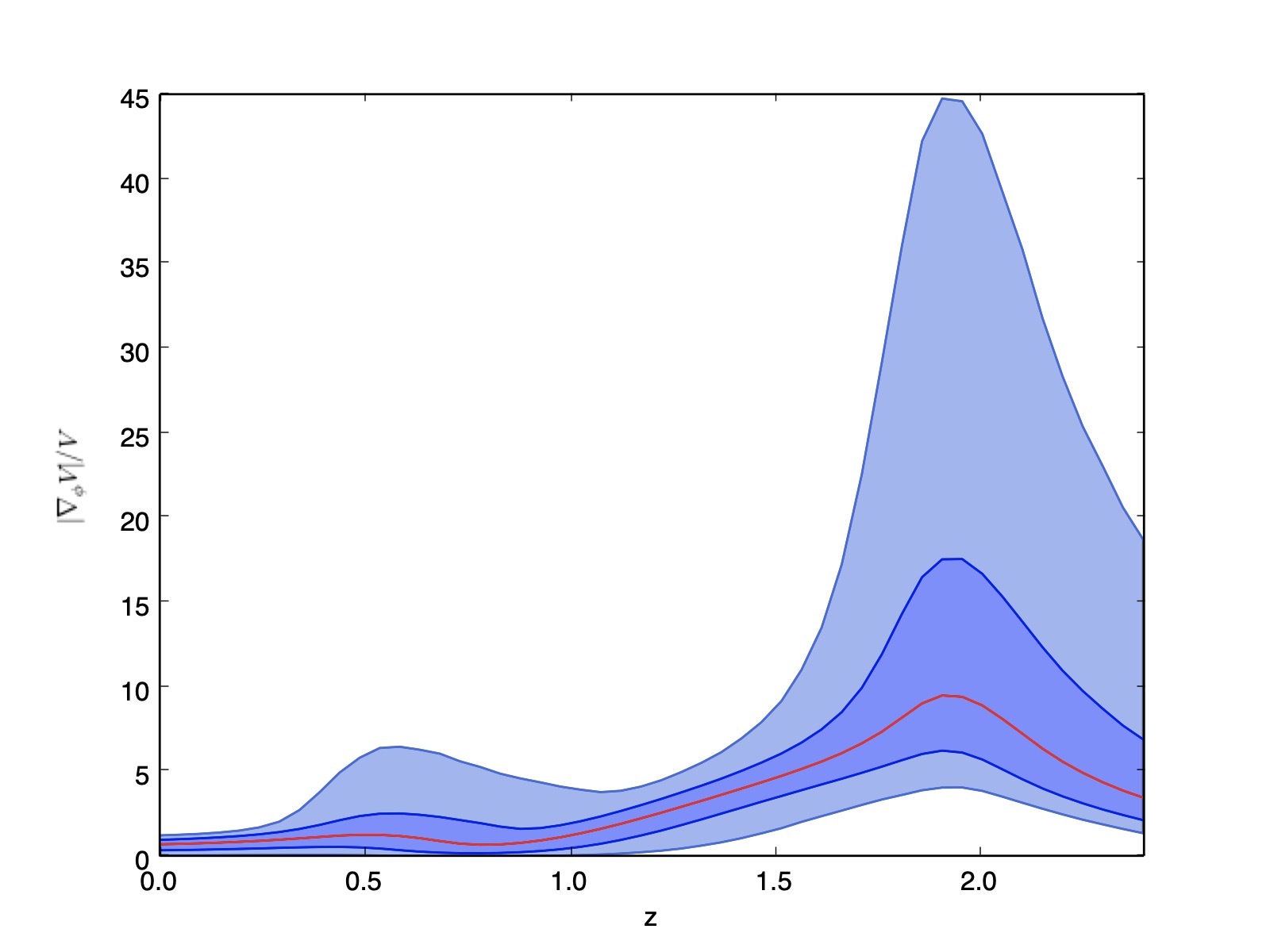} &&
\includegraphics[width=90 mm]{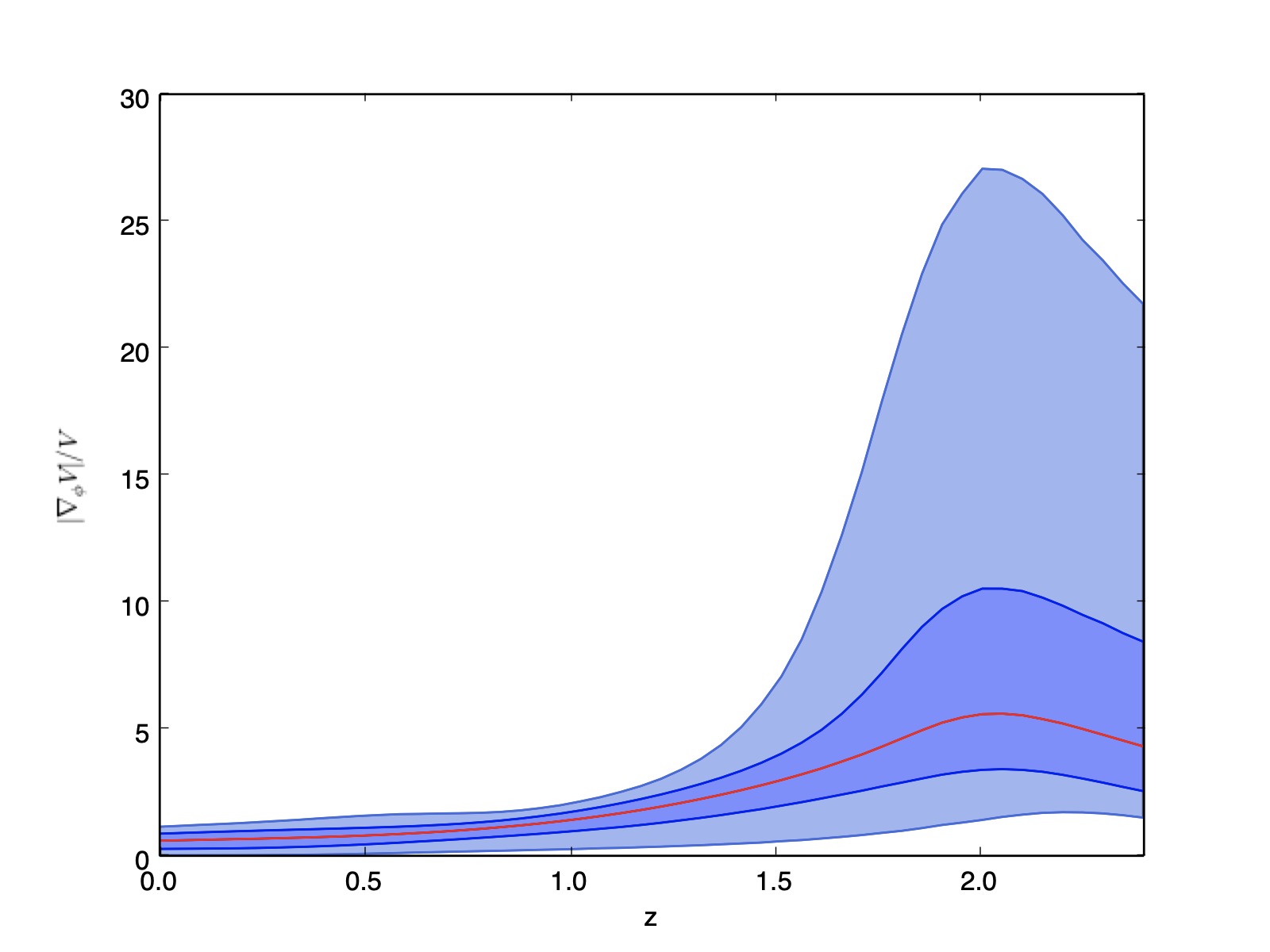} \\
 \end{array}$
 \end{center}
\caption{The left-hand side plot represents the reconstruction of the $|V^{\prime}|/V$, Eq.~(\ref{eq:SC2}), when the $H_{0} = 73.52 \pm 1.62$ km/s/Mpc from the Hubble Space Telescope has been merged with the $H(z)$ data depicted in Table~\ref{tab:Table0} and used in the reconstruction process. The right-hand side plot represents the reconstruction of the $|V^{\prime}|/V$, Eq.~(\ref{eq:SC2}), when the $H_{0} = 67.4 \pm 0.5$ km/s/Mpc from the Planck CMB data analysis has been merged with the $H(z)$ data depicted in Table~\ref{tab:Table0} and used in the reconstruction process. In both cases the squared exponent kernel given by Eq.~(\ref{eq:kernel1}) has been used. The solid line is the mean of the reconstruction and the shaded blue regions are the $68\%$ and $95\%$ C.L. of the reconstruction, respectively. The model is given by Eq. (\ref{eq:omega_2}) and Eq. (\ref{eq:NCDM_2}).}
 \label{fig:Fig2}
\end{figure}

The reconstruction of the Swampland criteria, Eq.~(\ref{eq:SC2}), when the squared exponent kernel is given by Eq.~(\ref{eq:kernel1}) can be found in Fig. \ref{fig:Fig2}. It clearly indicates that using only the Swampland criteria, Eq.~(\ref{eq:SC2}), a unique solution for the $H_{0}$ tension problem can not be found. Here we also have four possibilities indicating that, for instance, we can craft effective theories not residing in the Swampland region that can end up or not end up in Swampland at $z=0$. Besides this, we see that similar transitions are possible during the whole evolution of the universe where the deviation from the cold dark matter is given by Eq.~(\ref{eq:omega_2}) and Eq.~(\ref{eq:NCDM_2}). The constraints on $N_{eff}$ inferred from the Swampland criteria, expansion rate data, and GPs with three different kernels indicate that we should expect three generations of neutrinos in this universe. 

On the other hand, we learned that including neutrino strongly affects the constraints on the $\omega_{1}$ parameter. The last means that the study of the early universe can impose strong constraints on the dynamics in the deviations from the cold dark matter model. It could have a strong impact on various processes and the physics of the early universe requiring additional analysis that has been left to be tackled in the forthcoming papers. One of the initial results already shows that non-linear dynamics in the deviation can strongly affect the $N_{eff}$ indicating the existence of more than three neutrino generations in such models of the universe. 

To end the section we need to indicate that the cases we have analyzed allowed also to reproduce the $\Lambda$ model of dark energy. The reconstruction results indicated that phantom and quintessence universes can be obtained too. Moreover, the phantom divide from the above and the bottom can also be realized. We refer readers to \cite{Elizalde_H0_recent} where a detailed discussion of the behavior of the dark energy equation of state parameter $\omega_{de}$ recently appeared. To save our place we do not reproduce them here again.

\section{\large{Discussion}}\label{sec:Discussion}

The $H_{0}$ tension problem is one of the recent problems of modern cosmology attracting a lot of attention. Various attempts to solve it including solutions based on the interacting and early dark energy models already have been discussed and appeared in the literature. In this problem, it is required to explain why the Planck CMB data analysis and a local measurement from the Hubble Space Telescope give different values for $H_{0}$. We need to understand why in the $\Lambda$CDM scenario the Planck CMB data analysis gives $H_{0} = 67.4 \pm 0.5$ km/s/Mpc, while local measurements from the Hubble Space Telescope yield $H_{0} = 73.52 \pm 1.62$ km/s/Mpc. Nowadays even there is a huge hint that the problem challenges the $\Lambda$CDM model itself and it does not have an observational origin. Actually, the $\Lambda$CDM model has two components that over the years have been challenged. Indeed, various dark energy models already have been crafted and analyzed giving a good motivation to build modified GR theories, too. The second component is the cold dark matter and the challenge of it mainly started with the introduction of non-gravitational interaction between it and dark energy models. Because on the mathematical level, when non-gravitational interaction is introduced we modify the energy density dynamics of both components. Therefore in this scenario, the background dynamics will be described by some effective dark energy and dark matter models where the effective dark matter is not cold anymore. On the other hand, since observations do not exclude interacting dark energy scenarios, we seriously need to consider the possibility that on the cosmological scales, dark matter is not cold. Indeed, recently Bayesian machine learning a deviation from the cold dark matter has been learned. Moreover, the deviation has been confirmed using GP and expansion rate data. Detailed analysis showed that the stronger deviation from cold dark matter solves the $H_{0}$ tension problem. Some initial hints that the deviation has dynamic nature also has been revealed to be analyzed yet. On the other hand, Bayesian machine learning and GP already have been used to study the Swampland criteria in a dark energy-dominated universe. In particular, tension with the recent form of the Swampland criteria had been learned when the GP has been applied. In other words, we arrived at a point where using machine learning tools we have various interesting and promising results, but how they are connected to each other has not been analyzed yet. The goal of this work is to use GP and reveal how they are interconnected and what sort of new knowledge we can infer from them. 

Therefore, in this paper, we used GP to study the Swampland criteria for the scenarios where we have a deviation from the cold dark matter model. Moreover, we constrained the neutrino generation number and found that the tension with the recent form of the Swampland criteria still survives. This is another indication that interacting dark energy models and including neutrinos in analysis can not reduce the Swampland tension. We also found that considered cosmological models where the deviations from the cold dark matter are given by Eq. (\ref{eq:NCDM_1}) and  Eq. (\ref{eq:NCDM_2}), respectively, have three neutrino generations (perfect agreement with the predictions followed from GR). However, when the deviation has non-linear nature then there is a hint that more neutrino generations could exist. More detailed analyses of such scenarios have been left to be reported in the forthcoming papers. We also found that the Swamplang criteria alone do not offer a unique solution to the $H_{0}$ tension problem which confirms previously obtained results. Another important result capturing our attention is the constraints we have obtained on $\omega_{1}$ free parameter for the second model given by Eq. (\ref{eq:NCDM_2}). In particular, we learned that including neutrino significantly reduced the numerical value of this parameter indicating that the studies of the early universe will help us understand better what is the nature behind the deviations from the cold dark matter model. This is an important question requiring a serious study since with a deviation from cold dark matter our aim is to exclude non-gravitational interactions, however, we still need to understand how to do it correctly.  

To end we need to stress that in this paper only some steps are taken to collect the Swampland criteria, neutrino physics, the $H_{0}$ tension, and the deviation from the cold dark matter under the same umbrella using machine learning techniques. However, obtained results are very promising giving a hint that some fundamental problems of modern cosmology can be treated from a different point of view. Various questions still should be answered yet and their connections with learned results still should be established. Some of them, we have already mentioned and they have been left to be tackled in the forthcoming papers.

\section*{Acknowledgements}
M.K. has been supported by a Juan de la Cierva-incorporación grant (IJC2020-042690-I). This work has been partially supported by MICINN (Spain), project PID2019-104397GB-I00, of the Spanish State Research Agency program AEI/10.13039/501100011033, by the Catalan Government, AGAUR project 2021-SGR-00171, and by the program Unidad de Excelencia María de Maeztu CEX2020-001058-M.

\end{document}